\documentclass[conference]{IEEEtran}
\IEEEoverridecommandlockouts
\usepackage{cite}
\usepackage{amsmath,amssymb,amsfonts}
\usepackage{tikz}
\usepackage{aeguill}
\usepackage{graphicx}
\usepackage{textcomp}
\usepackage{xcolor}
\usepackage{color}

\usepackage{algpseudocode}
\usepackage{algorithm}

\algnewcommand\algorithmicforeach{\textbf{for each}}
\algdef{S}[FOR]{ForEach}[1]{\algorithmicforeach\ #1\ \algorithmicdo}

\def\BibTeX{{\rm B\kern-.05em{\sc i\kern-.025em b}\kern-.08em
    T\kern-.1667em\lower.7ex\hbox{E}\kern-.125emX}}
\begin{document}
\title{O-RAN for Energy-Efficient Serving Cluster Formulation in User-Centric Cell-Free MMIMO
\thanks{The presented work was funded by the Polish National Science Centre, project no. 2022/45/N/ST7/01930.}
}

\author{
\IEEEauthorblockN{Marcin Hoffmann}
\IEEEauthorblockA{\textit{Institute of Radiocommunications} \\
\textit{Poznan University of Technology}\\
Poznan, Poland
}
\and
\IEEEauthorblockN{Pawel Kryszkiewicz}
\IEEEauthorblockA{\textit{Institute of Radiocommunications} \\
\textit{Poznan University of Technology}\\
Poznan, Poland 
}
}


%


\maketitle

\begin{abstract}
The 6G Massive Multiple-Input Multiple-Output (MMIMO) networks can follow the so-called User-Centric Cell-Free (UCCF) architecture, where a single user is served by multiple Access Points (APs) coordinated by the Central Processing Unit (CPU). In this paper, we propose how O-RAN functionalities, i.e., rApp-xApp pair, can be used for energy-efficient Serving Cluster Formulation (SCF). Simulation studies show up to 37\% gain in Energy Efficiency (EE) of the proposed solution over the state-of-the-art Network-Centric (NC) designs.  
\end{abstract}


%
\IEEEpeerreviewmaketitle

\section{Introduction}

The UCCF MMIMO is an emerging technology for future 6G networks~\cite{Akyildiz2020}. Unlike the state-of-the-art NC approach, where a given user is typically served by one Base Station (BS), the UCCF MMIMO network is to provide the user with a coordinated transmission from multiple APs, that are coordinated by the CPU. This approach allows to significantly improve the network Spectral Efficiency (SE) and to provide users with equalized Quality of Service (QoS). Moreover, the UCCF network design can potentially improve the EE over the NC approach, as a result of the increased number of degrees of freedom. One of the key challenges in a UCCF MMIMO network is to determine which APs should serve a specific user, e.g., distant APs might provide a marginal contribution to user throughput while increasing signaling overhead, and "wasting" allocated power. The procedure of AP selection is known as the SCF. It has been shown that the practical implementation of the UCCF MMIMO network can follow the Open Radio Access Network (O-RAN) architecture~\cite{Ranjbar2022}. By utilizing the 7.2 split, the E2 Node (here O-RAN Distributed Unit) can serve as CPU and control transmission from multiple O-RAN Radio Units (O-RUs). The control of an O-RAN network is possible with the use of so-call xApps, and rApps that are deployed within Near Real-Time RAN intelligent Controller (Near-RT RIC), and Non-RT RIC, respectively. The former works within the control loop of time scale between 10~ms and 1~s while the latter above 1~s. While the SCF through the dedicated xApp has been proposed in~\cite{Beerten2022}, the authors did not consider the participation of Non-RT RIC and were focused on the SE and throughput, not EE. Moreover, the numerical analysis assumes a simple radio channel model and most importantly a narrowband, single-carrier system, while real 5G, and plausibly 6G, networks utilize Orthogonal Frequency Division Multiple Access (OFDMA). 

In this paper, we propose EE-oriented SCF that involves cooperation between rApp in Non-RT RIC and xApp in Near-RT RIC. The simulation studies are performed with the use of an advanced OFDMA-based UCCF MMIMO network simulator utilizing 3D Ray Tracing software.

\begin{figure}[!t]
\centering
\includegraphics[width=0.49\textwidth]{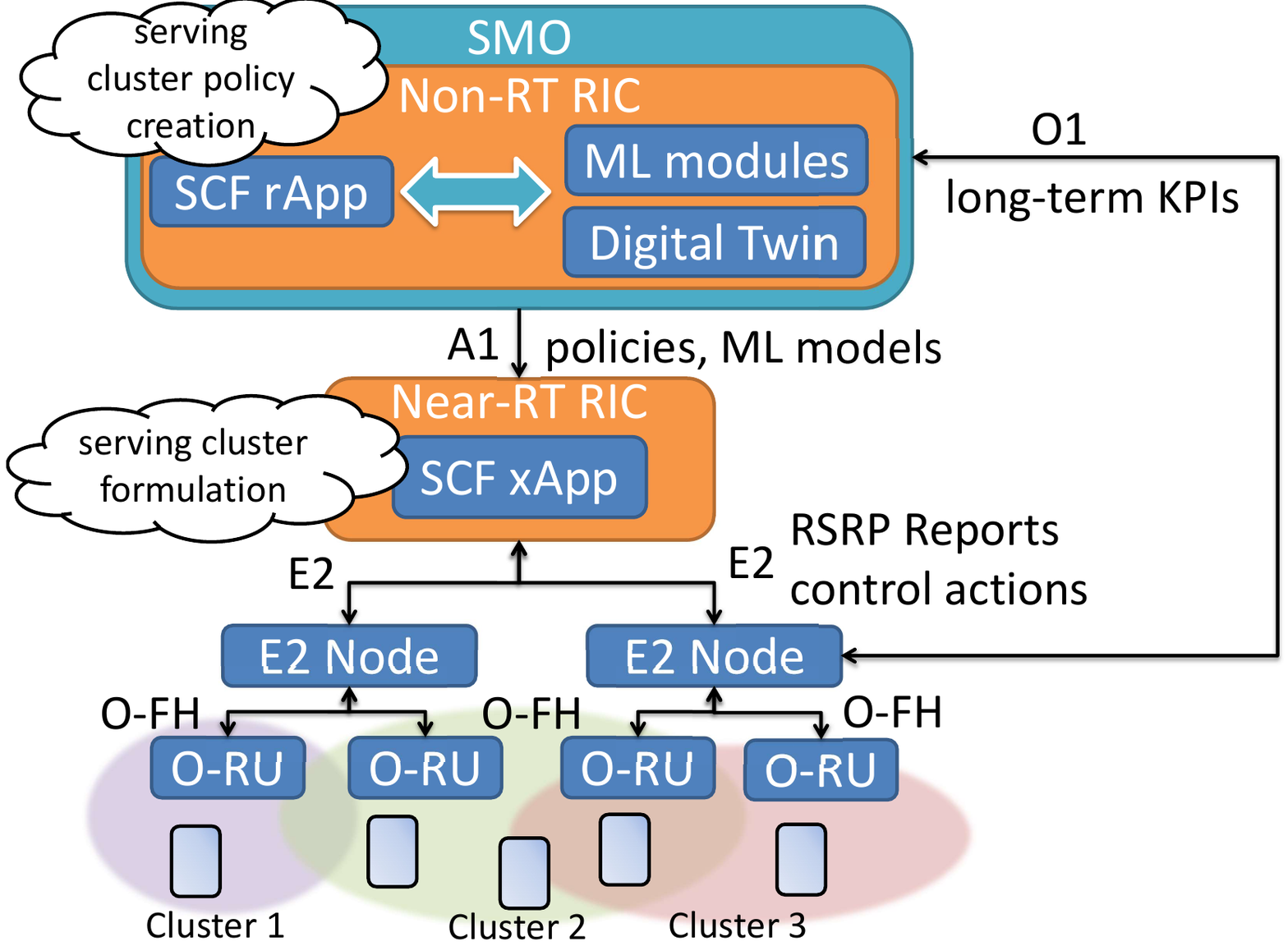}
\caption{The concept of rApp-xApp cooperation for EE-oriented SCF in O-RAN UCCF MMIMO network.}
\label{fig:cfxapp}
\end{figure}

\section{xApp/rApp for EE-oriented SCF}

The concept of rApp-xApp cooperation for EE-oriented SCF in O-RAN UCCF MMIMO network is depicted in Fig.~\ref{fig:cfxapp}. The SCF rApp placed in the Non-RT RIC is responsible for creating policies for SCF that contain, e.g., the number of O-RUs that should formulate a serving cluster, or SCF algorithm to be utilized, out of the set available in the xApp. These are communicated to the SCF xApp through the A1 interface, possibly together with pre-trained ML models. To formulate these policies SCF rApp utilizes 3GPP-compliant KPIs measured over the O1 interface~\cite{3gpp202328552}, e.g., UL/DL throughput distribution, and power consumption. Potentially rApp can also use information from ML models or Digtial Twin-based simulations deployed within the Non-RT RIC. On the other hand, the SCF xApp is responsible directly for SCF according to the A1 policies produced by rApp, and user Reference Signal Received Power (RSRP) measurements. Both control actions - SCF, and input data - RSRP reports are obtained by the E2 interface that connects Near-RT RIC with E2 Nodes (O-DUs). The E2 Nodes are further responsible for the configuration of O-RUs to create the desired serving clusters, e.g., synchronization of multi-point MMIMO transmission, channel estimation, and sharing of user data. However, enabling these features requires standardization effort both within the A1 and E2 interface, to enable new policy types and E2 Service Models (SMs) proper for the UCCF MMIMO network. 

To test the validity of the proposed approach an SCF based on the reported RSRPs is proposed. The serving cluster is formulated separately for each UE by selecting the $N$ O-RUs of the highest RSRPs. The $N$ is provided by the SCF rApp through A1 policy and refers to Serving Cluster Size (SCS). After the serving cluster is formulated the SCF rApp can monitor both per-user EE based on UL/DL throughput distribution and O-RU's power consumption, or RAN EE by dividing the total UL/DL throughput by the total power consumption. Such measurements can be used to train the ML model or to adjust Digital Twin (DT) to support SCF rApp with the formulation of policies that improve EE. 

\section{Simulation Results}

\begin{figure}[!t]
\centering
\includegraphics[trim={0.5cm 0.3cm 2cm 0.5cm},clip,width=0.49\textwidth]{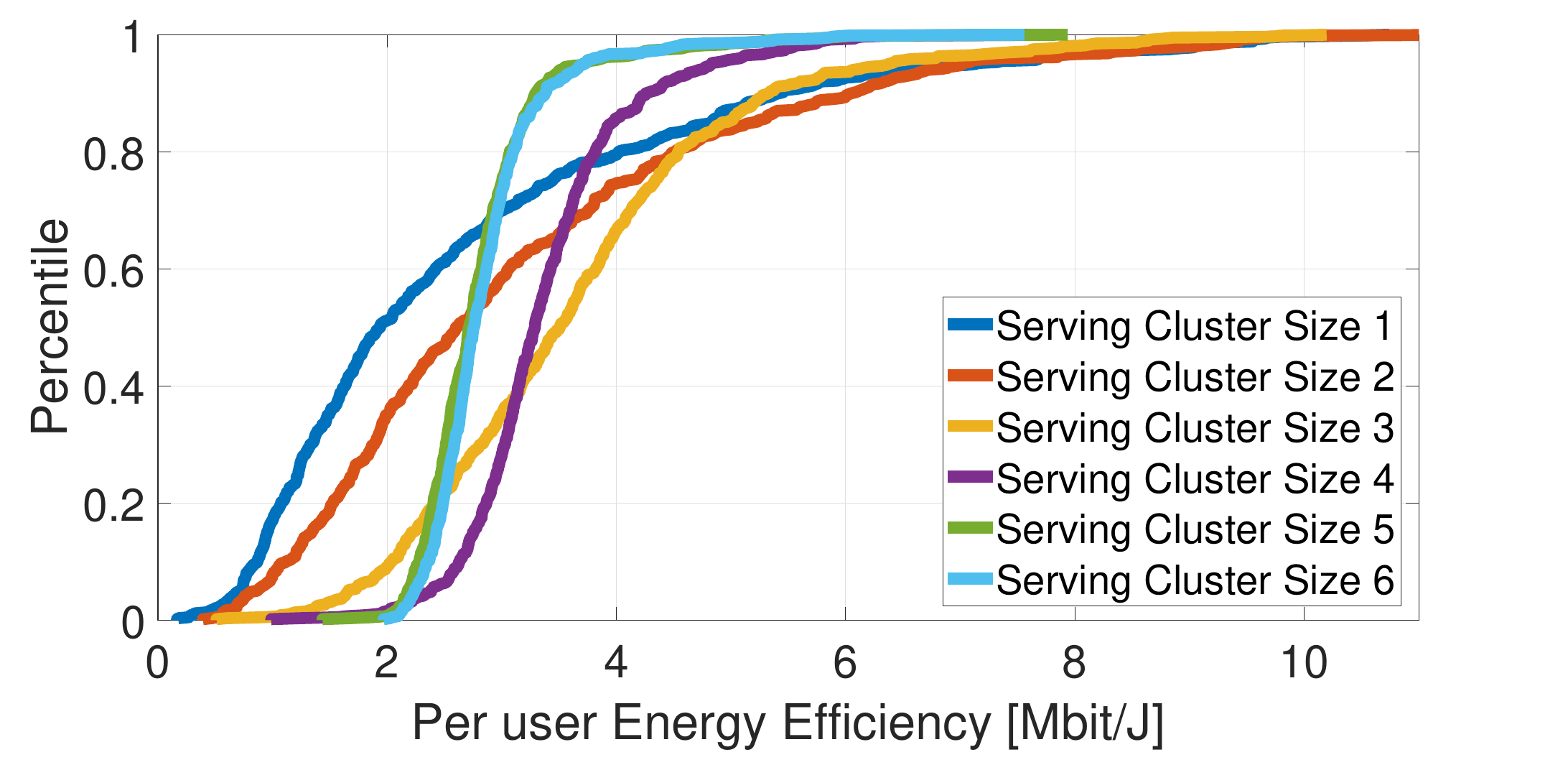}
\caption{Distribution of per-user EE for varying Serving Cluster Size.}
\label{fig:cdf_ue_ee}
\end{figure}

\begin{figure}[!t]
\centering
\includegraphics[trim={0.25cm 0.2cm 2cm 0.5cm},clip,width=0.49\textwidth]{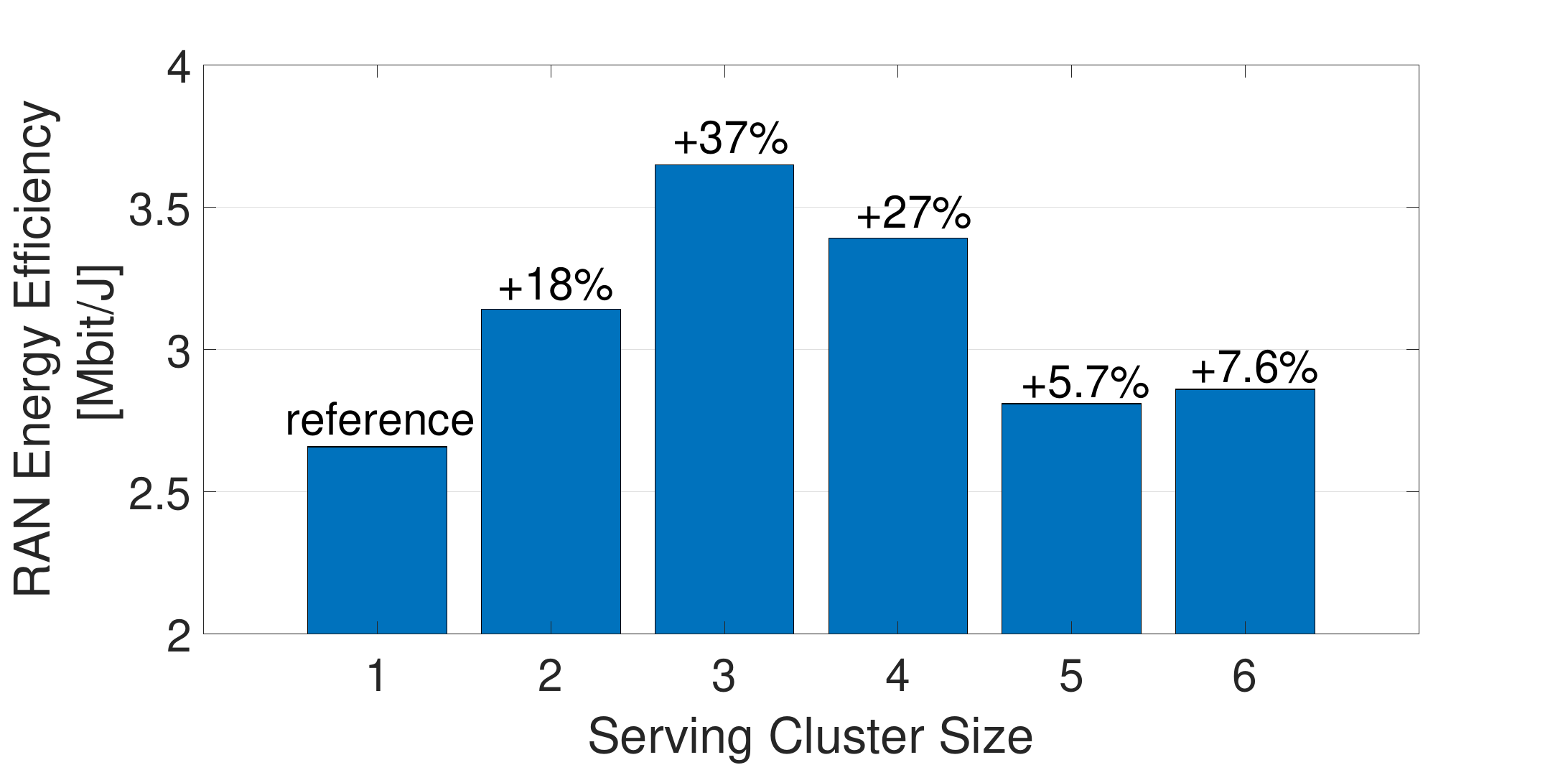}
\caption{RAN EE for varying serving cluster size.}
\label{fig:bar_sys_ee}
\end{figure}

The proposed xApp-rApp pair SCF algorithm is evaluated within an advanced network simulator of the OFDMA-based UCCF MMIMO network described in~\cite{hoffmann2024} that utilizes Wireless InSite 3D Ray Tracer, and dedicated radio resource scheduler. We assume that there is a single E2 Node responsible for radio resource scheduling that coordinates transmission from O-RUs. We consider 1 macro O-RU of transmit power equal to 46~dBm, equipped with 128 antennas, and 5 micro O-RUs of transmit power equal to 30~dBm, equipped with 32 antennas. The network utilizes 25~MHz bandwidth around a center frequency of 3.6~GHz. The Zero-Forcing (ZF) precoding is used independently within each O-RU. The results are obtained after 15 independent simulation runs, each emulating 500 ms of network operation, considering 20 randomly placed UEs. In Fig.~\ref{fig:cdf_ue_ee} a distribution of per-user EE is shown for varying SCS. The size equal to 1 corresponds to the NC approach. It can be seen that utilization of the UCCF approach provides per-user EE gains over the NC approach in at least $70\%$ of cases. Also for growing SCS the distribution of per-user EE becomes more uniform. However, for $\mathrm{SCS}>3$ EE of about 60\% of users starts to degrade compared to the $\mathrm{SCS}=3$.
A similar tendency can be observed in the RAN EE depicted in Fig.~\ref{fig:bar_sys_ee}. The highest EE is related to the $\mathrm{SCS}=3$. It provides gains of about 37\% over the NC approach. For higher SCS the EE gains degrade. 
The reason can be that while increasing the number of SCS too much, users are being served by some distant O-RUs. These O-RUs are "wasting" transmit power to overcome large path loss. 

\section{Conclusion}

We have shown the utilization of the proposed SCF xApp and rApp can provide significant EE gains in the O-RAN-based UCCF MMIMO network over the NC approach. However, to maximize gains, the SCS must be carefully chosen by SCF rApp based on the network state. In the future, this work can be extended by the utilization of DT or O-RU's micro sleep controlled by SCF xApp.

\bibliography{bibliography} 
\bibliographystyle{IEEEtran}

\end{document}